\documentclass[aps,prl,reprint,groupedadresses]{revtex4-1}
\usepackage{graphicx}
\usepackage[utf8]{inputenc}
\usepackage{amsmath}
\usepackage{amssymb}
\usepackage{color}
\renewcommand\vec[1]{\boldsymbol{#1}}

\newcommand\phihex{\ensuremath{\phi_{\text{hex}}}}
\newcommand\philiq{\ensuremath{\phi_{\text{\,liq}}}}
\newcommand\corrsix{\ensuremath{\xi_{\text{6}}}}
\newcommand\corrpos{\ensuremath{\xi_{\text{p}}}}
\newcommand\phisolid{\ensuremath{\phi_{\text{hs}}}}
\newcommand\Pcoex{\ensuremath{P_{\text{lh}}}}
\newcommand\ExPcoex{\ensuremath{\beta\Pcoex\sigma^2}}
\newcommand\rmi[2]{\ensuremath{{#1}_{\text{#2}}}}
\newcommand\kilo{\operatorname k}
\newcommand\Fig[1]{Fig.~\ref{#1}}
\newcommand\Tab[1]{Tab.~\ref{#1}}
\newcommand\coloronline{\emph{(Color online)}}

\date{\today}
\begin{document}

\title{2D Melting: From Liquid-Hexatic Coexistence
to Continuous Transitions}

\author{Sebastian C. Kapfer}
\email{sebastian.kapfer@fau.de}
\author{Werner Krauth}
\email{werner.krauth@ens.fr}
\affiliation{Laboratoire de Physique Statistique, Ecole Normale Sup\'{e}rieure,
UPMC, Universit\'{e} Paris Diderot, CNRS, 24 rue Lhomond, 75005 Paris, France}

\begin{abstract}
The phase diagram of two-dimensional continuous particle systems is
studied using Event-Chain Monte Carlo. For soft disks with repulsive
power-law interactions $\propto r^{-n}$ with $n \gtrsim 6$, the 
recently established hard-disk melting scenario ($n \to \infty$) holds: a
first-order liquid-hexatic and a continuous hexatic-solid transition are
identified. Close to $n = 6$, the coexisting liquid exhibits very long
orientational correlations, and positional correlations in the hexatic are
extremely short. For $n\lesssim 6$, the liquid-hexatic transition is continuous,
with correlations consistent with the Kosterlitz-Thouless-Halperin-Nelson-Yong (KTHNY) scenario. To illustrate the
generality of these results, we demonstrate that Yukawa particles likewise may follow
either the KTHNY or the hard-disk melting scenario, depending on the Debye-H\"uckel
screening length as well as on the temperature.
\end{abstract}

\maketitle 

Two-dimensional particle systems with short-range interactions may form 
solids \cite{Alder62}, but cannot acquire long-range positional
order \cite{WagnerMermin}.
Rather, two-dimensional solids are characterized by long-range
orientational and quasi-long-range positional order, so that positional
correlation functions decay algebraically. In the liquid phase, both orientational
and positional order are short-ranged, and the corresponding correlation
functions decay exponentially. An intermediate hexatic phase may also exist \cite{KTHNY}.
It is characterized by short-range positional and quasi-long-range
orientational order.

Within the Kosterlitz-Thouless-Halperin-Nelson-Young (KTHNY) theory of
two-dimensional melting \cite{KTHNY}, these two symmetry-breaking
transitions
arise from the subsequent unbinding of topological defects:  In the solid,
dislocations are bound in pairs, whereas in the hexatic, free dislocations may
exist. The dislocations then decompose into free disclinations which break
orientational order and yield the isotropic liquid. Both phase transitions are
of the continuous Kosterlitz-Thouless type, although a first-order
liquid-hexatic transition remains possible within the KTHNY framework.
Alternative theories of two-dimensional melting propose
a conventional first-order liquid-solid transition, in the absence of a
hexatic phase \cite{StrandburgReview,AlbaSimionescoReview}. These scenarios
commonly involve the condensation of defects into grain boundaries and related
aggregates \cite{Chui,KleinertBook,GeomCondens}.

Over the decades, it has been extremely difficult to decide, from theory,
simulation or experiments, which of the above melting scenarios applied to
specific two-dimensional models. It was established only very recently that
the fundamental hard-disk model has a continuous solid-hexatic transition 
but a first-order hexatic-liquid transition \cite{BernardKrauth}. This
scenario continues to apply for three-dimensional hard spheres tightly confined
between parallel plates \cite{Quasi2D}.  Indications for such a
scenario were also found for two-dimensional Yukawa particles \cite{QiYukawa}.

Experimentally, evidence for liquid-hexatic coexistence was reported both
for sterically stabilized uncharged colloids \cite{MarcusRice} and for charged
colloids \cite{LinChen}.  In complex plasmas, grain-boundary melting was reported
\cite{Plasmas}.  The KTHNY theory was confirmed experimentally for
superparamagnetic colloids \cite{Konstanz}. Other two-dimensional systems which
melt include electrons pinned at a liquid helium interface \cite{Helium}, and
surface-adsorbed atomic layers \cite{Xenon}.

\begin{figure}
\includegraphics{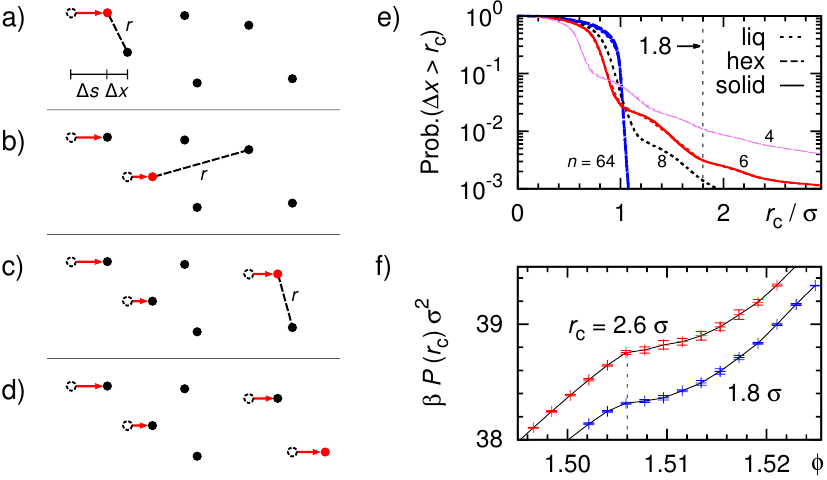}
\caption{\coloronline\ Event-Chain algorithm for continuous pair interactions.
a-d) Evolution
of the algorithm for four particles through three subsequent
collision events.
e) Probability for soft disks that a collision event takes place at a
distance $ r $ larger than the cutoff $\rmi{r}{c}$
The vertical dashed line is the cutoff chosen in this work.
f) Equation of state with different cutoffs ($N=65\kilo$, $n=6$).
\label{figECMC}}
\end{figure}

\begin{figure*}
\includegraphics{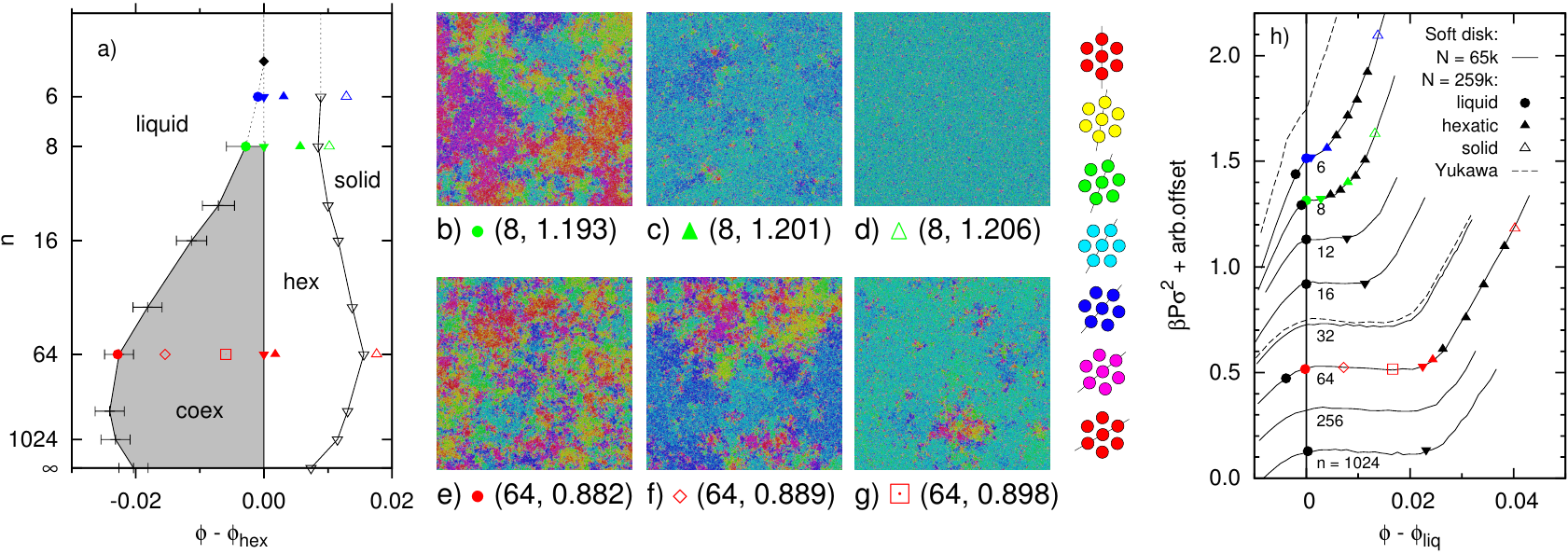}
\caption{\coloronline\ 
Phase behavior of $r^{-n}$ soft disks for $n \geq 6$. a) Phase diagram as a
function of density $\phi$ relative to the density \phihex\ of the pure hexatic at coexistence.
The non-monotonic liquid-hexatic coexistence interval vanishes around $n=6$.
Symbols match the following figures: bullets are liquid states, filled triangles hexatics,
of which downward filled triangles the hexatic at \phihex; empty triangles are solids.
Center: Local orientational order parameter $\psi_6$ in $N=259\kilo$ particles,
the color code is illustrated to the right.  Upper row:
b) Liquid phase at $(n, \phi) = (8, 1.193)$ (subset of a $1036\kilo$ configuration)
c) hexatic. d) solid.
Lower row: Coexistence in the $n=64$ system:  e) pure liquid close to coexistence.
f) At $\phi = 0.889$, the hexatic and the liquid form stripes.
g) At $\phi = 0.898$, a small bubble of liquid remains on a hexatic background
of uniform orientation.
h) Equations of state with $n$ from $6$ through $1024$ (solid lines correspond
to $N=65\kilo$ soft disks, symbols correspond to $N=259\kilo$, \philiq\ is
the liquid density at coexistence, dashed lines are for Yukawa particles).
\label{figALLES}
}
\end{figure*}

In this paper, we systematically study two-dimensional melting 
for repulsive pair interactions using computer simulations. We first concentrate
on the inverse power-law pair interaction $U(r) = \varepsilon (\sigma/r)^n$. 
This family of potentials includes hard disks
of diameter $\sigma$ (for $n\to\infty$) but also, at intermediate $n$, the soft interactions
typically found in colloidal particles, and long-range
interactions such as dipolar (for $n=3$, \cite{Konstanz}) and Coulomb forces
(for $n=1$, \cite{Helium}). We will establish that for large enough
$n$, the hard-disk melting scenario with its first-order liquid-hexatic
transition is preserved. Around $n = 6$, the system changes
over to the classical KTHNY scenario with two continuous transitions.  The hexatic
phase is firmly established for all parameters studied. To illustrate the
generality of our findings, we approximately map state points of soft disks
onto those of particle systems with Yukawa pair interactions by tuning the
Debye-H\"uckel screening length and the temperature.  We again identify both the
KTHNY and the hard-disk scenarios.

The soft-disk interaction sets no separate energy scale, and we may put
$\beta\varepsilon=1$, where $\beta = 1/\rmi kBT$
is the inverse temperature. The phase diagram only depends on the dimensionless
density $\phi=\sigma^2 N/V$, which is related to the dimensionless interaction
strength $\Gamma = \beta\varepsilon (\pi\phi)^{n/2}$. Length scales can be
expressed in terms of  the interparticle distance $d = (\pi N/V)^{-1/2}$. In
these units, the pair interaction is $\beta U = \Gamma \times (d/r)^n$.

To accommodate the large correlation lengths inherent in two-dimensional melting,
we consider systems  of $65\cdot 10^3$ ($65 \kilo$), $259\kilo$ and $1036\kilo$ particles.
To equilibrate these large systems, we use
Event-Chain Monte Carlo (ECMC) \cite{BernardKrauth}, recently parallelized
\cite{ParallelEC} and extended to continuous interactions \cite{GeneralEC}.
ECMC displaces a single active particle (red bullet in \Fig{figECMC}a) in
a fixed direction in successive infinitesimal steps corresponding to
a continuous Monte Carlo time. Instead of rejecting a move
because of a potential barrier 
between two particles (dashed line), the displacement is transferred from one
particle to the other
(\Fig{figECMC}b). In this way, cooperative cluster moves are built up.  The
algorithm is most easily understood for hard-sphere systems, but the concept
of pairwise collision events carries over to continuous interactions.
ECMC samples the canonical ensemble exactly \cite{GeneralEC}. It is implemented
efficiently using an event-driven approach and mixes faster than conventional
Monte Carlo \cite{GeneralEC}. Truncating
the interaction $\tilde U(r) = U(\min(r, \rmi rc))$ at a cutoff distance $\rmi rc$
amounts to neglecting some collision events \cite{Endnote}.  For soft disks with
$n\geq 6$, using the cutoff $\rmi rc =  1.8\sigma$, less than $2\cdot 10^{-3}$
of collision events are missed (\Fig{figECMC}e). ECMC yields the pressure as a
zero-cost byproduct of the simulation \cite{GeneralEC}, and allows for the
construction of the equation of state.  For the cutoff $\rmi{r}{c} = 1.8\sigma$,
the pressure is reduced by about $1.1\%$ with respect to the usual cutoff
$2.6\sigma$, but the phase boundaries (e.~g., the liquid-hexatic kink in
\Fig{figECMC}{f} at $\phi=1.506$) are not moved, as we checked
explicitly.

For each of the $n$ studied, we find extended liquid, hexatic, and solid
phases (see \Fig{figALLES}{a--d}).  In the liquid, both positional correlations
and the correlations of the local orientational order parameter $\psi_6$ (with
Voronoi weights \cite{Mickel}) are short-ranged,  the latter is visualized in
\Fig{figALLES}{b, e}.  For large $n$, the equation of state displays
a clear Mayer-Wood loop \cite{MayerWood} characteristic
of a first-order transition (see \Fig{figALLES}{h}; cf.\ Ref.~\cite{BernardKrauth} for a discussion of
phase coexistence in the $NVT$ ensemble). At $n=64$, the liquid-hexatic
coexistence interval is wider in density than for hard disks ($n= \infty$), yet
qualitatively equivalent: As in Ref.~\cite{BernardKrauth}, we observe both stripe-shaped
coexisting phases (see \Fig{figALLES}{f}), and bubble-shaped minority phases
(\Fig{figALLES}{g}). For smaller $n$, the coexistence interval narrows
(see \Fig{figALLES}{a, h}) and finally vanishes around $n=6$, where the
transition becomes continuous.  The computational cost of long-ranged interactions
due to larger cutoffs makes large-scale simulations for $n<6$ prohibitively slow.

\begin{figure}
\includegraphics{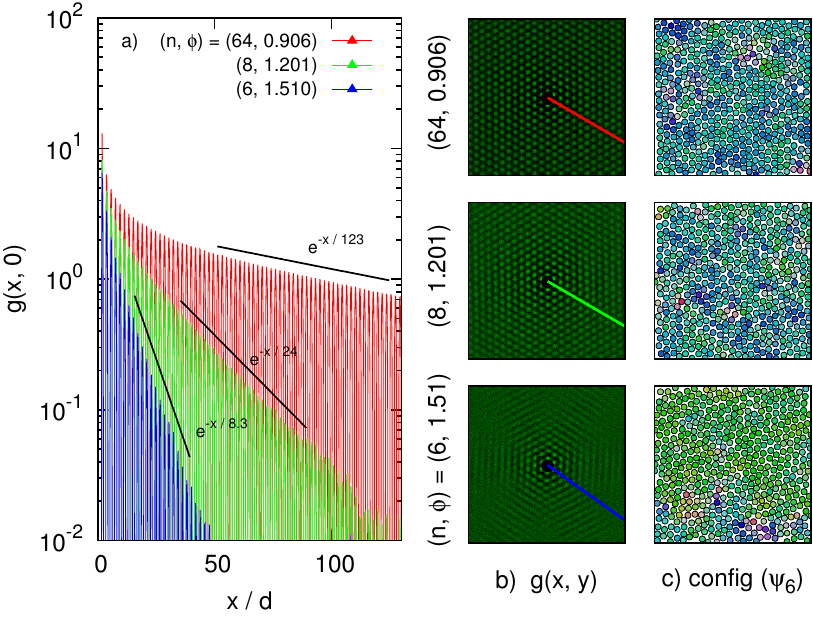}
\caption{\coloronline\ a)
Pair correlation function $g(x,y)$ along
the $x$ axis in the pure hexatic phase, showing exponential decay for large $x/d$
(ensemble average of $N=259\kilo$ configurations after aligning their global
orientational order parameters $\Psi_6$, as in Ref.~\cite{BernardKrauth}).
b) Two-dimensional pair correlation function $g(x,y)$ for single configurations
at the same parameters. The solid line is the $x$ axis for the left-hand plot.
c) Square boxes of side length $40 d$ extracted from the $259\kilo$ configurations:
Orientational order is preserved as the positional order is lost (color code
for $\psi_6$ as in \Fig{figALLES}{b--g}).
\label{figPositionalSRO}}
\end{figure}

\begin{figure}
\includegraphics{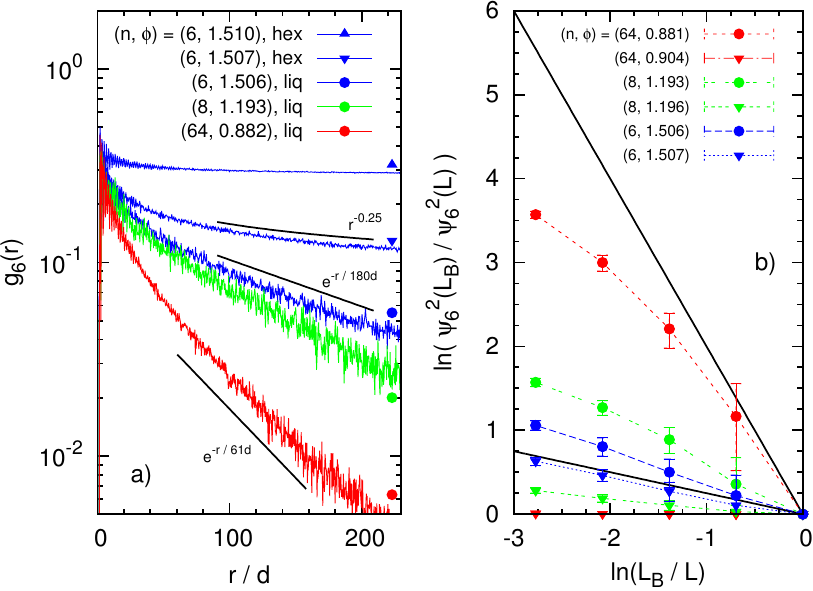}

\caption{\coloronline\ a) Orientational correlations $g_6(r)$
close to the liquid-hexatic transition, for several exponents $n$. 
b) Scaling of the orientational order parameter in subblocks of
linear size $L_\text B$ (see Ref.~\cite{BagchiAndersen} for details).
The KTHNY hexatic is stable below the bold line of
slope $-\frac 14$; short-range order corresponds to the steep bold line of slope $-2$.
Bullets are liquid states, triangles hexatics (symbols match those in \Fig{figALLES}).
\label{figSix}}
\end{figure}

The phase in coexistence with the liquid is a hexatic. Since correlation functions are
ambiguous in the coexistence region, we consider pure hexatics above the transition, and
find short-ranged positional correlations (\Fig{figPositionalSRO}{a}) while
orientational correlations are quasi-long-ranged (see \Fig{figSix}{a}).  The
positional correlation length \corrpos\ can be as large as $100$ interparticle
distances $d$, but we can equilibrate systems of sufficient size to reveal the
asymptotic exponential decay of the ensemble-averaged pair correlation function
$g(r)\propto\exp(-r/\corrpos)$.  In the lowest-density pure hexatic, at the transition,
\corrpos\ decreases strongly with $n$ as the coexistence interval vanishes (see
\Tab{tabData}).  At $n=6$, the correlation length is on the order
of the interparticle distance $d$.  Single-configuration pair correlations also
confirm that positional order drops from $\corrpos\approx 100d$ at $n=64$ to a
few neighbors at $n=6$ (see \Fig{figPositionalSRO}{b}).

We now turn to the analysis of orientational correlations in the
transition region.  We extract the corresponding correlation length
\corrsix\ in the liquid from the asymptotic exponential decay of the correlation
function $g_6(r)=\langle\psi_6(\vec r_i)\psi_6(\vec r_j)\delta(r-r_{ij})\rangle$.
The orientational correlation length is large but finite
in the pure liquid at coexistence. It increases markedly from
$n=64$ to $n=8$ and  $n=6$ (see the lower three curves in \Fig{figSix}{a}, data in
\Tab{tabData}).

At $n=6$, a minute increase in density changes the orientational order from 
short-range to an algebraic decay with exponent
$\approx -\frac 14$ (\Fig{figSix}{a}).
This agrees with the KTHNY prediction of  orientational
correlations $\propto r^{\eta}$ in the hexatic, with $\eta=-\frac{1}{4}$ at the
transition.
Away from the transition point, and for $n>6$, the orientational correlation function
does not display clear power-law behavior.  This is also borne out by the
finite-size scaling technique of Ref.~\cite{BagchiAndersen}, computing
the average orientational order $\Psi_6(L_\text B) = \langle \psi_6 \rangle_\text B$
in subblocks of linear size $L_\text B$.  Due to finite-size effects, the liquids at
coexistence deviate from the ideal short-range behavior (steep bold line of slope $-2$
in \Fig{figSix}{b}), but they are well beyond the KTHNY stability limit for the hexatic
(bold line of slope $-\frac 14$).  The $n=64$ and $n=8$ hexatics at the transition have
small slopes (\Fig{figSix}{b}, $\eta=-0.0026$ and $-0.10$), while for $n=6$, we find
a value close to the stability limit, $\eta=-0.19$.
Our data are thus consistent with a continuous Kosterlitz-Thouless transition for $n\lesssim 6$,
which is preempted by a first-order transition for larger $n$.

Approaching the hexatic-solid transition, $\corrpos$ increases, and the ECMC
algorithm falls out of equilibrium. Moreover, the effective lattice constant,
reduced by a finite
equilibrium concentration of defects, is a priori unknown and the positional
order in our samples is usually incommensurate with the periodic boundary conditions,
leading to frustration effects.  This prevents robust conclusions for the exact
density of the hexatic-solid transition density \phisolid. Nevertheless, 
we can provide a lower bound for the melting density \phisolid\ from the
highest-density configurations that could be
molten in a $N=259\kilo$ periodic box (\Tab{tabData}).  At the hexatic-solid transition, we find
no indications of a discontinuity; in particular, the equation of state shows
no Mayer-Wood loop (\Fig{figALLES}{h}).

\begin{figure}
\includegraphics{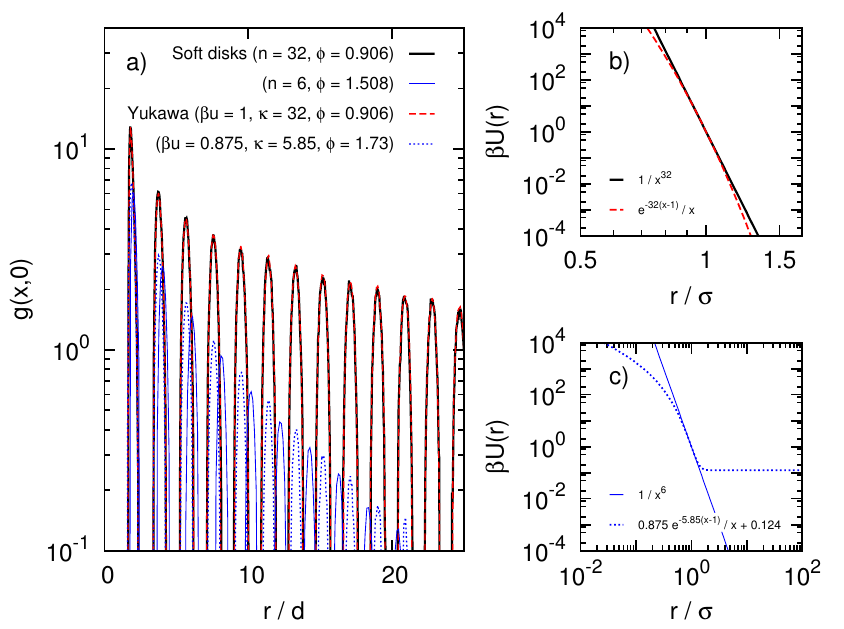}
\caption{\coloronline\ Yukawa interactions exhibiting first-order (top)
and continuous liquid-hexatic transitions.  a) Pair correlation function
along the $x$ axis, as in \Fig{figPositionalSRO}a.
b, c) Interaction potential of the soft disks and the Yukawa particles around
$r=\sigma$.
\label{figYukawa}}
\end{figure}

While the correlation lengths in the hexatic are so vastly different in the two
regimes, we observe no qualitative difference in the structure of the KTHNY 
topological defects, which we define as coordination anomalies in the Voronoi diagram
\cite{StrandburgReview}, even though this type of analysis is not without
problems in disordered systems \cite{Mickel}.  The behavior of defects is
collective more so than the simple subsequent unbinding picture would suggest.
While free dislocations (i.\,e., isolated pairs of a 5- and a 7-coordinated
site) exist and free disclinations are indeed heavily suppressed, the majority
of defects are involved in more complex clusters.  Most defects form stringlike
aggregates in which topological charges alternate, and frequently comprise an
odd number of bound dislocations.  Their classification into grain boundaries,
disclinations and dislocations becomes ambiguous.
The defect strings circumscribe patches of homogeneous orientational order, and
the liquid-hexatic transition can be viewed as the percolation transition of the
defect string network.  Thus, the liquid-hexatic transition occurs according to
a grain-boundary mechanism, similar to proposed direct liquid-solid transitions
\cite{Chui,KleinertBook}, but starting from a hexatic phase.

The change of scenario from liquid-hexatic coexistence to a continuous transition 
occurs not only in soft disks as a function of $n$, but whenever the interaction
forces can be tuned
between the hard-disk and long-range limits.  Effectively, only a small part of the
interaction potential is explored by the particles:  For potentials
that, in the relevant range of interparticle distances,
are well approximated by a soft-disk potential, the soft-disk phase behavior
should be recovered.  To test this hypothesis, we use the
Yukawa interaction $U(r) = u \times (\sigma /r) \exp (\kappa(1-r/\sigma))$, and
match its first and second derivatives at $r=\sigma$ to the respective derivatives
of the soft-disk interaction by tuning the effective interaction strength
$\beta u$ and the Debye-H\"uckel screening length $\kappa^{-1}$ (see \Fig{figYukawa}b, c).
Indeed, we find for the Yukawa system corresponding to soft disks with $n=32$
(parameters $\beta u=1$,
$\kappa=32$) a first-order transition (see the lower dashed curve in
\Fig{figALLES}h) into a hexatic with long positional correlation length
$\corrpos\sim 100 d$ (\Fig{figYukawa}a).  Approximating
the $n=6$ soft-sphere interaction on the other hand ($\beta u=0.875$, $\kappa=5.85$),
the transition is clearly continuous (upper dashed curve in \Fig{figALLES}h)
and the hexatic phase has extremely short positional correlation lengths
(\Fig{figYukawa}a).  These findings agree qualitatively with the results for soft disks.
Thus, the change of scenarios identified in this work should be observable experimentally,
for example, in charged colloids, planar plasmas, etc., by tuning the Debye-H\"uckel
screening length.

We have shown in this work that two-dimensional melting in particle systems 
with short-range repulsive pair interactions is generically a two-step
transition, with a hexatic phase
between the liquid and the solid.  We identify  two regimes:  At large $n$, and
for strong screening in the Yukawa particles, we recover the hard-disk melting scenario.
In the hexatic phase, at large $n$, positional correlation
lengths are two orders of magnitude larger than the interparticle distance $d$.
The density of positional defects is correspondingly small. As the interaction
potential becomes softer, the nature of the hexatic changes: positional correlation
lengths drop to a few $d$, and defects are ubiquitous.  The additional entropy
due to defects stabilizes the hexatic phase with respect to the liquid state, and
the phase-coexistence interval becomes very small.
For even smaller $n \lesssim 6$, the liquid-hexatic transition turns
continuous, and we recover the standard KTHNY scenario.
Conversely, it appears also possible to shift the liquid-hexatic
first-order transition towards \emph{higher} densities.  In this case, the
hexatic region and the hexatic-solid transition can be preempted by the
first-order transition, giving rise to a direct liquid-solid transition.  This
has been reported for the ``core-softened'' potential \cite{Coresoftened} which
includes (at low temperatures) a potential shoulder destabilizing hexagonal
order and favoring a direct solid-liquid transition at high density. For large
$T$, the core-softened potential of Ref. \cite{Coresoftened} reduces to the
$r^{-14}$ interaction considered here, and would consequently follow a two-step
melting scenario with the intermediate hexatic phase.
Owing to the long-range nature of interactions, the regime of
extremely soft and long-ranged potentials, $n<6$, is not presently accessible
to our large-scale simulations, but no further change of scenario is expected
for even smaller $n$.  Such systems have been considered in experiment
\cite{Konstanz}, and we expect that the change from classical KTHNY
two-step melting to a first-order liquid-hexatic transition followed by a
Kosterlitz-Thouless-type hexatic-solid transition can be tested experimentally.

This work was supported by a grant of computer time at the PSL Computing
Centre MesoPSL.  We thank H.~Kleinert, O.~Dauchot, and the referees for
insightful comments.

\begin{table}
{\renewcommand\tabcolsep{.5em}
\begin{tabular}{l|lllll|l}
$n$    & \ExPcoex  & \philiq &$\corrsix/d$& \phihex &$\corrpos/d$& \phisolid \\
6      &  38.3     &  1.506  &  180       &  1.507  &  2.6       & $>1.516$ \\
8      &  23.1     &  1.193  &  110       &  1.196  &  6.0       & $>1.204$ \\
12     &  14.7     &  0.998  &  112       &  1.005  &  13        & $>1.015$ \\
16     &  12.1     &  0.937  &  95        &  0.949  &  27        & $>0.960$ \\
64     &  9.27     &  0.882  &  61        &  0.904  &  96        & $>0.920$ \\
1024   &  9.17     &  0.889  &  65        &  0.913  &  66        & $>0.924$ \\
$\infty$& 9.18     &  0.892  &  62        &  0.913  &  51        & $>0.919$
\end{tabular}}
\caption{Thermodynamic data for $r^{-n}$ soft disks:
Pressure $\Pcoex$ at the liquid-hexatic transition;
density \philiq\ and orientational correlation length \corrsix\ of the liquid;
density \phihex\ and positional correlation length \corrpos\ of the hexatic at
coexistence.  The final column is a lower bound for the melting density (solid-hexatic).
The densities are accurate to  $\approx 0.5\%$.  Pressures are computed
using the truncated interaction and are thus low by at most $1.1\%$.
The statistical sampling error is a decade smaller (error bars in \Fig{figECMC}f).
The correlation lengths are determined from the tail of the respective correlation
functions and are subject to large errors of $\pm 10\%$ due to the choice of $\phi$.
They are consistent with earlier results for hard disks \cite{BernardKrauth}.
\label{tabData}
}
\end{table}

\end{document}